# Variance-based uncertainty relation for incompatible observers

Xiao Zheng(郑晓) ,Guo-Feng Zhang(张国锋)[1,*]

*Key Laboratory of Micro-Nano Measurement-Manipulation and Physics (Ministry of Education), School of Physics and Nuclear Energy Engineering, Beihang University, Xueyuan Road No. 37, Beijing 100191, China*

**Abstract:** Based on mixedness definition as $M = 1 - \text{tr}(\rho^2)$, we obtain a new variance-based uncertainty equality along with an inequality for Hermitian operators of a single-qubit system. The obtained uncertainty equality can be used as a measure of the system mixedness. A qubit system with feedback control is also exploited to demonstrate the new uncertainty.

**I. Introduction**

Uncertainty relation lies at the heart of quantum mechanics, providing one of the most fundamental departures from classical concepts [1-15]. Any pair of incompatible observables admits a certain form of uncertainty relationship, and this constraint sets the ultimate bound on the measurement precision achievable for these quantities. As originally formulated by Heisenberg, the conventional variance-based uncertainty relation not only possesses a clear physical conception, but also has a variety of applications in quantum information science such as entanglement detection [5, 16, 17], quantum spin squeezing [18-21], and even quantum metrology [22-24]. Among them Robertson uncertainty relation (RUR) is the most famous one [2]:

$$(\Delta A)^2 (\Delta B)^2 \geq \left| \frac{1}{2i} \langle [A, B] \rangle \right|^2, \qquad (1)$$

where the standard deviation $\Delta O$ and expectation value $\langle O \rangle$ are taken over the state $\rho$ with $O \in \{A, B\}$. A strengthened form of RUR is due to Schrodinger [3, 25-27], who derived the following Schrödinger uncertainty relation (SUR):

$$(\Delta A)^2 (\Delta B)^2 \geq \left| \frac{1}{2i} \langle [A, B] \rangle \right|^2 + \left| \frac{1}{2} \langle \{\check{A}, \check{B}\} \rangle \right|^2, \qquad (2)$$

where $I$ denotes the identical operator and $\check{O} = O - \langle O \rangle I$. Uncertainty inequalities similar to Eq.

---

(1) and Eq. (2) are often referred to as Heisenberg-type and Schrödinger-type uncertainty relation, respectively. There are also many different ways to quantify the measurement uncertainty, such as Entropic uncertainty relation (EUR) [8-14] and Heisenberg's Noise-disturbance uncertainty [28-31]. The initial version of EUR is given by Kraus [4]:

$$H(A) + H(B) \geq \log_2 \frac{1}{c}, \qquad (3)$$

where $H(A)$ is the Shannon entropy of the probability distribution of the outcome when $A$ is measured, and likewise for $H(B)$. $\log_2(1/c)$ quantifies the complementarity of $A$ and $B$, where $c = \max_{a,b} |\langle \Psi_a | \Phi_b \rangle|^2$ with $|\Psi_a\rangle$ and $|\Phi_b\rangle$ being the eigenvectors $A$ and $B$. A modification of the entropic uncertainty relation occurs in the presence of quantum memory associated with quantum correlations [6]. The uncertainty relations for multi-observable have also been formulated besides two-observable uncertainty relations [27].

The outline of the paper is as follows. In Sec. II, we obtain a new variance-based uncertainty equality along with an inequality for Hermitian operators of a single-qubit system on the basis of SUR and RUR. The obtained uncertainty equality can be used as a measure of system mixedness which usually is expensive in terms of resources and measurements involved. In Sec. III, We show using the example of a qubit system with feedback that the tightness of the uncertainty inequality can be maintained at a high level even in an open system. Finally, Sec. IV is devoted to the discussion and conclusion.

**II. Deduction of the new variance-based uncertainty equality**

The mixedness definition of the system, the construction of the new variance-based uncertainty equality, and the relevant discussion are presented in this section.

*Definition of Mixedness*: A state $\rho$ is a mixed one when $0 < \text{tr}(\rho^2) < 1$ and $\text{tr}(\rho^2) = 1$ for the pure one. Therefore, the value of $M$ can be employed to quantify the system mixedness by denoting $M = 1 - \text{tr}(\rho^2)$. The convexity of mixedness is derived as (for more detail please see Appendix):

$$M_{(x\rho_A + (1-x)\rho_B)} \geq x M_{(\rho_A)} + (1-x) M_{(\rho_B)}, \qquad (4)$$

where $\rho_A$ and $\rho_B$ represent two arbitrary density matrices, $x\rho_A + (1-x)\rho_B$ is the combination of them with $0 \leq x \leq 1$ and $M_{(\rho)}$ stands for the mixedness of the state $\rho$ ($\rho \in \{\rho_A, \rho_B, x\rho_A + (1-x)\rho_B\}$).

*Variance-based Uncertainty equality*: Let $A$ and $B$ stand for two arbitrary Hermitian operators of the single-qubit system, then the uncertainty of the outcomes when they are measured admit the following equality:

$$(\Delta A)^2(\Delta B)^2 = \left|\frac{1}{2i}\langle[A,B]\rangle\right|^2 + \left|\frac{1}{2}\langle\{\check{A},\check{B}\}\rangle\right|^2 + \frac{1}{8}M[\xi(A,A)\xi(B,B) - \xi(A,B)^2], \quad (5)$$

where $\xi(R,S) = 2\text{tr}(RS) - \text{tr}(R)\text{tr}(S)$ with $R, S \in \{A, B\}$.

*Proof*: In Bloch sphere representation, the density matrix of the single-qubit system can be expressed as [25]:

$$\rho = \frac{1}{2}(I + p_x\sigma_x + p_y\sigma_y + p_z\sigma_z), \quad (6)$$

where $I$ stands for the identity matrix, $\sigma_x, \sigma_y, \sigma_z$ are standard Pauli matrices and $p_x, p_y, p_z$ denote real parameters with $p_x^2 + p_y^2 + p_z^2 \leq 1$. After simple calculation, the system mixedness is obtained as:

$$M = \frac{1}{2}(1 - p_x^2 - p_y^2 - p_z^2). \quad (7)$$

It is well known that an two-dimension Hermitian operator can be written as a linear combination of $\{\sigma_x, \sigma_y, \sigma_z, I\}$:

$$A = a_1\sigma_x + a_2\sigma_y + a_3\sigma_z + a_4 I. \quad (8)$$

$$B = b_1\sigma_x + b_2\sigma_y + b_3\sigma_z + b_4 I. \quad (9)$$

where $a_i$ and $b_i$ are real parameters $(i = 1,2,3,4)$. Based on the above, we have:

$$(\Delta A)^2 = (1 - p_x^2)a_1^2 + (1 - p_y^2)a_2^2 + (1 - p_z^2)a_3^2 - 2[p_y p_z a_2 a_3 + p_x a_1(p_y a_2 + p_z a_3)], \quad (10)$$

$$(\Delta B)^2 = (1 - p_x^2)b_1^2 + (1 - p_y^2)b_2^2 + (1 - p_z^2)b_3^2 - 2[p_y p_z b_2 b_3 + p_x b_1(p_y b_2 + p_z b_3)], \quad (11)$$

$$\left|\frac{1}{2i}\langle[A,B]\rangle\right|^2 = (p_x(a_3 b_2 - a_2 b_3) + p_y(a_1 b_3 - a_3 b_1) + p_z(a_2 b_1 - a_1 b_2))^2, \quad (12)$$

$$\left|\frac{1}{2}\langle\{\check{A},\check{B}\}\rangle\right|^2 = ((p_x^2 - 1)a_1 + p_x(p_y a_2 + p_z a_3))b_1 + (p_x p_y a_1 + (p_y^2 - 1)a_2 + p_y p_z a_3)b_2 + (p_x p_z a_1 + p_y p_z a_2 + (p_z^2 - 1)a_3)b_3, \quad (13)$$

$$\text{tr}(A^2) = 2a_1^2 + 2a_2^2 + 2a_3^2 + 2a_4^2, \quad (14)$$

$$\text{tr}(B^2) = 2b_1^2 + 2b_2^2 + 2b_3^2 + 2b_4^2, \quad (15)$$

$$\text{tr}(A) = 2a_4, \ \text{tr}(B) = 2b_4, \quad (16)$$

$$\text{tr}(AB) = 2a_1 b_1 + 2a_2 b_2 + 2a_3 b_3 + 2a_4 b_4. \quad (17)$$

The uncertainty equality can be obtained by some simple substitution and then the proof is completed.

It is notable that the contribution of the mixedness is mainly reflected in the remainder of the new equality and the uncertainty degenerates into SUR when the system mixedness reaches minimum value. In addition, the obtained uncertainty equality can be used as a measure of system mixedness which usually is expensive in terms of the resource and measurement involved. The expression of the mixedness can be obtained by a deformation of Eq. (5):

$$M = \frac{8[(\Delta A)^2(\Delta B)^2 - \left|\frac{1}{2i}\langle[A,B]\rangle\right|^2 - \left|\frac{1}{2}\langle\{A,B\}\rangle - \langle A\rangle\langle B\rangle\right|^2]}{\xi(A,A)\xi(B,B) - \xi(A,B)^2} \quad (18)$$

According to Eq. (18), the system mixedness can be easily obtained by detecting the variance and expectation involved. Finally, a new Heisenberg-type uncertainty inequality is derived by using the non-negativity of $\left|\langle\{\tilde{A}, \tilde{B}\}\rangle\right|^2$,

$$(\Delta A)^2(\Delta B)^2 \geq \left|\frac{1}{2i}\langle[A,B]\rangle\right|^2 + \frac{1}{8}M[\xi(A,A)\xi(B,B) - \xi(A,B)^2], \quad (19)$$

The superiority of Eq. (19) will be demonstrated in section III.

### III. Qubit system with feedback as an illustration

Qubit system, a most widely used physical platform in the quantum information processing, has played an irreplaceable role in theoretical analysis and experimental tests. It would be of great interest to investigate the performance of the new uncertainty relations and compare them with other forms in the context of qubit systems. Quantum feedback control [32], which manipulates the system based on the information acquired by the measurement of the controlled system, will be added into the qubit system to control the system evolution more effectively. As shown in Fig.1, the physical model of a single atom resonantly coupled to a cavity with feedback will be introduced.

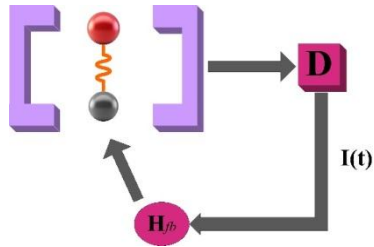

Fig. 1: Schematic view of the model, the feedback Hamiltonian is applied to the atoms according to the homodyne current I(t) derived from detector D, The coupling strength between the atom and the cavity is g, the two levels of the qubit are |0⟩ and |1⟩, and the spontaneous decay of the atom is γ.

An effective damping rate $\mathcal{T} = g^2/\kappa$ can be acquired under the condition that the cavity decay $\kappa$ is much larger than the other relevant frequencies of the system [33]. In the limit $\mathcal{T} \gg \gamma$, the spontaneous emission of the atoms is neglected and the dynamical evolution of this system can be described by the Dick model [33-36]:

$$\frac{d\rho}{dt} = -i[\Omega\sigma_x, \rho] + D(\mathcal{T}\sigma_-)\rho \qquad (20)$$

where $\rho$ is the density matrix of the qubit, $\Omega$ is the Rabi frequency, $\Omega\sigma_x$ represents the driving of the laser field, $\sigma_- = |0\rangle\langle 1|$ is the lowering operators of the qubit and the super operator $D(\mathcal{O})\rho = \mathcal{O}\rho\mathcal{O}^+ - (\mathcal{O}^+\mathcal{O}\rho + \rho\mathcal{O}^+\mathcal{O})/2$ represents the irreversible evolution induced by the interaction between the system and environment. In the following we will take $\mathcal{T} = 1$ to simplify the calculation.

The effect of feedback on the system will then be taken into consideration. As shown in Fig. 1, the output of the cavity is measured by a detector, and the signal $I(t)$ from the detector triggers a time-continuous feedback Hamiltonian. The master equation takes the form [35,36]

$$\frac{d\rho}{dt} = -i\left[\Omega\sigma_x + \frac{1}{2}(\sigma_+ F + F\sigma_-), \rho\right] + D(\sigma_- - iF)\rho \qquad (21)$$

Under the condition (i) $\Omega = 0$, namely taking no laser driving into consideration; (ii) $|\varphi(t = 0)\rangle = \cos(\alpha)|0\rangle + \sin(\alpha)|1\rangle$; (iii) $F = \lambda\sigma_x$ with $\lambda \in [0,1]$, the evolved density matrix of the qubit can be exactly solved:

$$\rho(t) = \begin{pmatrix} \rho_{11}(t) & \rho_{12}(t) \\ \rho_{12}^*(t) & 1 - \rho_{11}(t) \end{pmatrix}, \qquad (22)$$

with the elements

$$\rho_{11}(t) = \frac{e^{-t(1+2\lambda^2)}[1 + 2e^{t+2t\lambda^2}\lambda^2 - (1+2\lambda^2)\cos(2\alpha)]}{2(1+2\lambda^2)}, \qquad (23)$$

$$\rho_{12}(t) = \frac{e^{-t/2}(-i + ie^{-2t\lambda^2} + \lambda)\sin(2\alpha)}{2\lambda}. \qquad (24)$$

Since the uncertainty equality (5) is saturated for any single-qubit states, we focus on the performance and the superiority of the Heisenberg-type inequality (19). In Ref. [26] another uncertainty is derived:

$$(\Delta A)^2 + (\Delta B)^2 \geq \frac{1}{2}[\Delta(A + B)]^2 \qquad (25)$$

The comparison between Eq. (3), Eq. (19) and Eq. (25) will be presented in the following. It turns out to be a relatively reasonable way to compare different types of uncertainty relations through dividing both sides of the inequalities by their own lower bound. Therefore, the tightness of them

is defined as:

$$\text{Ti}_1 = \frac{(\Delta A)^2(\Delta B)^2}{\left|\frac{1}{2i}\langle[A,B]\rangle\right|^2 - \frac{1}{8}M[\xi(A,A)\xi(B,B)-\xi(A,B)^2]} \quad (26)$$

$$\text{Ti}_2 = \frac{H(A)+H(B)}{\log_2\frac{1}{c}} \quad (27)$$

$$\text{Ti}_3 = \frac{(\Delta A)^2+(\Delta B)^2}{\frac{1}{2}[\Delta(A+B)]^2} \quad (28)$$

The analytic expressions of $\text{Ti}_1$, $\text{Ti}_2$ and $\text{Ti}_3$ can be obtained under the condition $A = \sigma_x$, $B = \sigma_z$ and $\lambda = 1$.

$$\text{Ti}_1 = 1 + \frac{[-1+e^{3t}+3\cos(2\alpha)]^2\sin(2\alpha)^2}{8e^{7t}-e^t(1-3\cos(2\alpha))^2-2e^{4t}(3\cos(2\alpha)-1)-9e^{6t}\sin(2\alpha)^2} \quad (29)$$

We will not give the expressions of $\text{Ti}_2$ and $\text{Ti}_3$ due to their complicated form. The evolutions of $\text{Ti}_1$, $\text{Ti}_2$ and $\text{Ti}_3$ with respect to time and the initial state are shown in Fig.2:

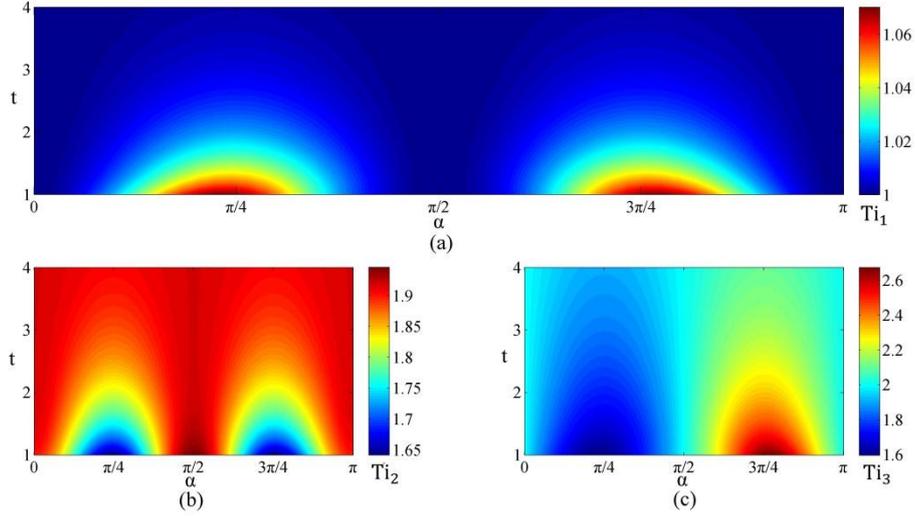

Fig. 2: the evolution of the tightness $\text{Ti}_i$ with $(\alpha, t)$, $\text{Ti}_1$ in (a), $\text{Ti}_2$ in (b) and $\text{Ti}_3$ in (c), here we take $\lambda = 1$.

It can be seen from the Fig.2 that the new uncertainty is tighter than the other two forms. Choosing $|\varphi_0\rangle = (|0\rangle + |1\rangle)/\sqrt{2}$ as the initial state leads to the following expression:

$$\text{Ti}_1 = \frac{(1-e^{-t})(1+2e^{t+2t\lambda^2}\lambda^2)(2e^{t+2t\lambda^2}(1+\lambda^2)-1)}{e^{t+2t\lambda^2}[2+e^{2t\lambda^2}(4\lambda^2(e^t-1)(\lambda^2+1)-1)]-1} \quad (29)$$

The evolution of $\text{Ti}_1$ with respect to $(t, \lambda)$ is illustrated by Fig. 3

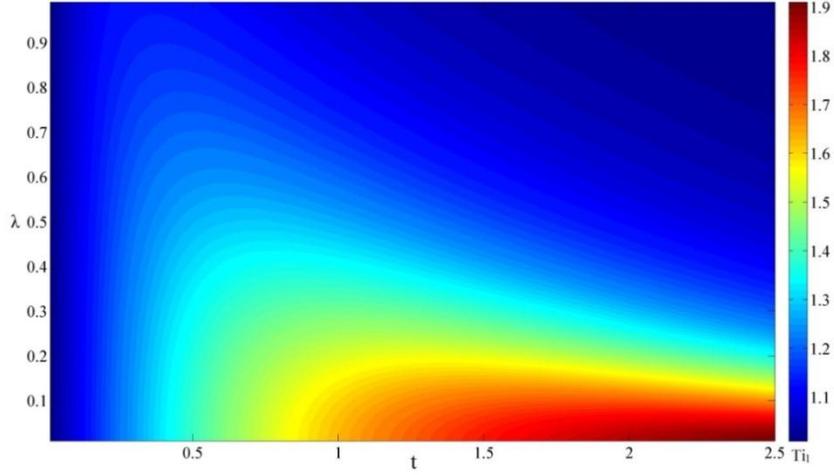

Fig. 3: the evolution of tightness $Ti_1$ with $(\lambda, t)$

As shown in Fig.3, the tightness of the new Heisenberg-type inequality can be maintained at a high level with time evolution by adjusting the feedback control, even in an open system, which is a very useful and meaningful resource in quantum information processing.

### IV. Conclusions

In conclusion, we construct and formulate a variance-based Schrodinger-type uncertain equality along with a Heisenberg-type uncertainty inequality, which hold for all pairs of incompatible observables of the single-qubit system. The obtained equality can be used as a measure of system mixedness. As an illustration, the qubit system with a feedback is investigated to demonstrate the superiority of the new uncertainty inequality. It is found out that the tightness of the new inequality can be maintained at a higher level with time evolution, even in open system.

### Acknowledgments

This work is supported by the National Natural Science Foundation of China (Grant No. 11574022).

## Appendix

With the inner product and the linearly independent basic vector of $d \times d$ dimension density matrix space being:

$$\langle \rho, \sigma \rangle = \text{tr}(\rho^+ \sigma). \tag{A1}$$

$$\{I,\ \Pi_1, \Pi_2, \dots \Pi_{d^2-1}\}, \tag{A2}$$

where $I$ stands for the identical matrix, and $\text{tr}(\Pi_i) = 0$, the density matrixes and their combination in the Bloch representation can be written as

$$\rho_A = \frac{1}{d}\left(I + \sum_{i=1}^{d^2-1} p_i^A \Pi_i\right), \tag{A3}$$

$$\rho_B = \frac{1}{d}\left(I + \sum_{i=1}^{d^2-1} p_i^B \Pi_i\right), \tag{A4}$$

$$x\rho_A + (1-x)\rho_B = \frac{1}{d}\{I + \sum_{i=1}^{d^2-1}[xp_i^A + (1-x)p_i^B]\Pi_i\}, \tag{A5}$$

where $0 \leq x \leq 1$ and $0 \leq \sum_{i=1}^{d^2-1} p_i^{\Theta^2} \leq 1$ ($\Theta \in \{A, B\}$). Based on the above,

$$\text{tr}(\rho_A^2) = \text{tr}(\rho_A^+ \rho_A) = \frac{1}{d^2}\left(d + d\sum_{i=1}^{d^2-1} p_i^{A^2}\right), \tag{A6}$$

$$\text{tr}(\rho_B^2) = \text{tr}(\rho_B^+ \rho_B) = \frac{1}{d^2}\left(d + d\sum_{i=1}^{d^2-1} p_i^{B^2}\right), \tag{A7}$$

$$\text{tr}([x\rho_A + (1-x)\rho_B]^2) = \frac{1}{d^2}\left(d + d\sum_{i=1}^{d^2-1}[xp_i^A + (1-x)p_i^B]^2\right), \tag{A8}$$

can be obtained, and then we have

$$\text{tr}([x\rho_A + (1-x)\rho_B]^2) \leq x\text{tr}(\rho_A^2) + (1-x)\text{tr}(\rho_B^2), \tag{A9}$$

where the inequalities $p_i^{A^2} + p_i^{B^2} \geq 2p_i^A p_i^B$ has been used. Thus, the convexity of system mixedness is derived as

$$M_{(x\rho_A + (1-x)\rho_B)} \geq xM_{(\rho_A)} + (1-x)M_{(\rho_B)}, \tag{A10}$$